\documentclass[]{acmart}
\usepackage{tabularx}

\AtBeginDocument{%
  \providecommand\BibTeX{{%
    \normalfont B\kern-0.5em{\scshape i\kern-0.25em b}\kern-0.8em\TeX}}}

\setcopyright{acmcopyright}
\copyrightyear{2021}
\acmYear{2021}
\acmDOI{10.1145/1122445.1122456}

\acmConference[ACM FAccT 2022]{ACM Conference on Fairness, Accountability, and Transparency 2022}{21-24 June, 2022}{Seoul, South Korea}
\acmBooktitle{ACM Conference on Fairness, Accountability, and Transparency 2022}
\acmPrice{15.00}
\acmISBN{978-1-4503-XXXX-X/18/06}

\begin{document}

\title{A Framework for Deprecating Datasets: Standardizing Documentation, Identification, and Communication}

% The \author macro works with any number of authors. There are two commands
% used to separate the names and addresses of multiple authors: \And and \AND.
%
% Using \And between authors leaves it to LaTeX to determine where to break the
% lines. Using \AND forces a line break at that point. So, if LaTeX puts 3 of 4
% authors names on the first line, and the last on the second line, try using
% \AND instead of \And before the third author name.

\author{Alexandra Sasha Luccioni}
\email{sasha.luccioni@huggingface.co}
\affiliation{
  \institution{Hugging Face}
  \country{Canada}
 } 
 
\author{Frances Corry}
\email{corry@usc.edu}
\affiliation{
  \institution{University of Southern California}
  \country{USA} }

\author{Hamsini Sridharan}
\email{hamsinis@usc.edu}
\affiliation{
  \institution{University of Southern California}
  \country{USA}
}

\author{Mike Ananny}
\email{ananny@usc.edu}
\affiliation{
  \institution{University of Southern California}
  \country{USA}
}

\author{Jason Schultz}
\email{SchultzJ@exchange.law.nyu.edu}
\affiliation{
  \institution{New York University}
  \country{USA} 
}

\author{Kate Crawford}
\email{kate.crawford@usc.edu}
\affiliation{
  \institution{University of Southern California, Microsoft Research}
  \country{USA}
}

\renewcommand{\shortauthors}{Luccioni and Corry et al.}
\renewcommand{\shorttitle}{A Framework for Dataset Deprecation}

\begin{abstract}
Datasets are central to training machine learning (ML) models. The ML community has recently made significant improvements to data stewardship and documentation practices across the model development life cycle. However, the act of deprecating, or deleting, datasets has been largely overlooked, and there are currently no standardized approaches for structuring this stage of the dataset life cycle. In this paper, we study the practice of dataset deprecation in ML, identify several cases of datasets that continued to circulate despite having been deprecated, and describe the different technical, legal, ethical, and organizational issues raised by such continuations. We then propose a Dataset Deprecation Framework that includes considerations of risk, mitigation of impact, appeal mechanisms, timeline, post-deprecation protocols, and publication checks that can be adapted and implemented by the ML community. Finally, we propose creating a centralized, sustainable repository system for archiving datasets, tracking dataset modifications or deprecations, and facilitating practices of care and stewardship that can be integrated into research and publication processes.
\end{abstract}



\maketitle

\section{Introduction}

Datasets are at the heart of machine learning (ML) theory and practice, driving progress in model development and serving as benchmarks for measuring this progress. A growing wave of scholarly work has focused on studying datasets, their contributions towards the success of ML models, and how they are created and maintained~\cite{gebru2018datasheets,mitchell2019model, crawford2019excavating,harvey2019exposing,peng2021mitigating}. However, the process of \textit{dataset deprecation}, i.e. the end-of-life phase in the model life cycle, has received far less attention despite the downstream effects deprecation practices can have on ML models. Datasets can be deprecated for many reasons, ranging from issues around legality (e.g., datasets with images gathered without proper consent) and ethics (e.g., datasets that perpetuate harmful stereotypes or biases), but also more mundane reasons like the creation of a new dataset version or the end of a grant. In all deprecation cases, to ensure datasets' integrity and legitimacy, the ML community needs practices and mechanisms to ensure that  deprecations are appropriate, well documented, clearly communicated, and follow a timeline and due process that allow for appeals. Without these mechanisms, ad-hoc deprecation practices can and have become the norm, resulting in datasets that continue to circulate after deprecation, either intentionally or unintentionally. When these datasets are cited in academic papers or used for training and deploying ML models, the problems compound. In some cases, multiple versions and derivatives of datasets live on after deprecation and continue to do harm, despite their creators’ intentions.  

We approach the problem of dataset deprecation with the good faith understanding that researchers continue to use these sites not out of malevolent intent but instead from limited discussion and oversight within the ML research community, making it extremely difficult to know which datasets have been deprecated and why. The use of deprecated datasets is further sustained by the lack of documentation about when, why, and how a dataset (or a part thereof) was removed. Poor documentation practices around dataset deprecation, along with their continued circulation and use within ML communities, can perpetuate the harms that deprecations often seek to address. 
While this framework focuses specifically on the deprecation of datasets within ML contexts, its themes relate to other fields that have likewise grappled with the ethics of information removal. For decades now, medical and scientific fields have contended with questions around if, when and how to retract journal articles that were based on shoddy data or drew false conclusions. Strategies developed for journal article retraction offer generative parallels for thinking through dataset deprecation. For example, discussions around journal article retraction have prompted the use of watermarks and other notices to signal an article’s retracted status, while professional bodies have been formed to specifically tackle these issues, like the Committee on Publication Ethics (COPE) (Nair et al 2019; Newman 2010).

This is our motivation for proposing the Dataset Deprecation Framework, which we hope will help researchers and practitioners track the status of datasets and reasons for their deprecation. Over time, standardized use of the framework may show the ML community patterns in dataset deprecation, revealing which motivations and rationales drive dataset deprecations and hopefully spurring open and critical conversation about the role that datasets play in model building. In addition, regular use of the framework may reveal best practices for dataset deprecation that are not yet evident given the lack of field-level discourse. We start, in Section~\ref{related}, with a brief overview of related work on dataset documentation, maintenance and care, from within the ML community and from other disciplines such as Science, Technology, and Society (STS). In Section ~\ref{outline}, we share a retrospective analysis of several popular ML datasets that have been deprecated, examining the reasons given for their deprecations and the actions taken by their creators. We identify some commonalities among the cases, including a lack of transparency and communication around the reasons for deprecation, as well as a lack of centralized directory for a dataset's status. We continue with a breakdown of issues that arise when datasets are not properly deprecated, ranging from the technical to legal, in Section~\ref{problem}. Finally, in Section~\ref{framework}, we propose a Dataset Deprecation Framework with three elements: a deprecation report, a system of permanent identifiers, and a centralized repository for managing deprecation. We conclude with a discussion of how our proposed framework can help to ensure proper dataset deprecation, supported by individual contributors and organizations within the ML community.

\section{Related work on dataset stewardship, maintenance, and care} \label{related}

Datasets are an important part of machine learning systems, serving both as training data for creating and deploying ML models as well as allowing researchers to benchmark the progress made on tasks such as image recognition and natural language understanding. Despite their importance, work on improving the quality and documentation of datasets has been largely under-valued by the community, with the greater focus given to developing better models and approaches~\cite{sambasivan2021everyone}. However, this focus has started to shift in recent years, during which a flurry of ML scholarship has unearthed many issues with even the most commonplace datasets, showing them to be riddled with errors, perpetuating bias, and intensifying discrimination~\cite{northcutt2021,crawford2019excavating, harvey2019exposing, prabhu2020large,luccioni2021s}. This has brought attention to the importance of conscientious data stewardship practices in all steps of the dataset life cycle, ranging from proper documentation at the data creation stage via checklists~\cite{madaio2020co} and data sheets~\cite{gebru2018datasheets} to the study of dataset usage and genealogy~\cite{koch2021reduced, denton2021genealogy}, licensing and auditing~\cite{benjamin2019towards,brown2021algorithm, raji2020saving, sandvig2014auditing}, as well as data stewardship and maintenance~\cite{paullada2020data, peng2021mitigating,yang2020towards}. 

There has also been extensive data-related scholarship among STS researchers focused on care and responsibility for sociotechnical systems, including \textit{why} to care for them, \textit{who} cares for them, and, crucially, \textit{how} to care for these systems, and proposing ways to make such practices part of sociotechnical cultures~\cite{martin_politics_2015, puig_de_la_bellacasa_matters_2011}. More concretely, safeguarding data and metadata through time has been studied through both social and technical relations~\cite{paskin2010digital} and STS scholars have shown how work like maintenance and repair (versus system construction and innovation) are often overlooked parts of understanding the ongoing functioning of technical systems~\cite{vinsel2020innovation}. This work has shown how values persist in sociotechnical systems in unacknowledged ways, past creation points and into the practices that maintain, repair, and dismantle what was initially constructed. In drawing attention to processes that recognize finitude and to forms of work that are unrecognized or undervalued within dominant cultures of innovation, this scholarship shows how sociotechnical maintenance and repair align and intertwine with  feminist political philosophies of care (e.g. ~\cite{cowan1983more}).

The maintenance of datasets over time can thus be understood as a care practice. A dataset’s lifecycle – its creation, use, deprecation, continued circulation – is a value-inflected social process that depends significantly upon the motivations, decisions, and actions of dataset creators. Data management is a significant challenge for the field, and there is now an urgent need for transparently motivated, clearly articulated, community-wide practices for dataset governance and stewardship. Thus our focus here is to support dataset creators and the wider ML community with a framework for maintaining and caring for a dataset at the end of its life. Applying this last question to the deprecation of training data, this paper does not ask those who interact with datasets to simply “care more,” but rather provides a scaffolding through which care practice can be enacted at a dataset’s end of life. In the following section, in order to illustrate the general lack of deprecation practices in our community, we study several major ML datasets and their deprecation, identifying commonalities and shortcomings that show the need for a common framework to guide and structure deprecation, given the potential harms and errors that can come from deprecated datasets' continued circulation and use.

%
%Care work has been theorized over decades of feminist scholarship, notably within studies of science, technology, and society~\cite{linden_editorial_2021, puig_de_la_bellacasa_matters_2011}.  As these studies have argued, care is “a practice, an acting, a doing”~\cite{martin_politics_2015}, an active form of thoughtful attention emphasizing interconnectedness, finitude, and vulnerability.

\section{A Study of Dataset Deprecation in Machine Learning} \label{outline}

A retrospective analysis of the deprecations of common ML datasets is important because it can show the strengths and weaknesses of different deprecation practices. In Table~\ref{table1}, we analyze the deprecation process of several major datasets: ImageNet, Tiny Images, MS-Celeb-1M, Duke MTMC, MegaFace, and HRT Transgender, noting external critiques of the deprecations, reasons given for the deprecations, and actions taken by the dataset creators.

\begin{table*}[h!]
\vspace*{-0.2cm}%
\begin{tabular}{@{}llll@{}}
\toprule
\textbf{Dataset} &
  \textbf{External critique} &
  \textbf{Reason given for deprecation} &
  \textbf{Actions taken} \\ \midrule
\multicolumn{1}{l|}{ImageNet} &
  \multicolumn{1}{l|}{\begin{tabular}[c]{@{}l@{}}Excavating AI project \\ publishes a critique of \\ ImageNet's problematic \\ “Person” categories in 2019 \\\cite{crawford2019excavating, metz2019nerd}.\end{tabular}} &
  \multicolumn{1}{l|}{\begin{tabular}[c]{@{}l@{}}Yang et al.~\cite{yang2020towards} confirm that of \\ 2,832 subcategories in “Person” \\ subtree, 1,593 are  “potentially \\  offensive.” Of the remaining \\ 1,239 categories, “only 158 are \\ visual” and should be filtered.\end{tabular}} &
  \begin{tabular}[c]{@{}l@{}}In 2019 over 600,000 images \\ associated with “potentially \\ offensive” categories removed \\ from ImageNet~\cite{ruiz2019leading,yang2020towards}. \\ In 2021, following Yang et al.~\cite{yang_study_2021},  \\ people’s faces are blurred when \\ they  appear “incidentally.”\end{tabular}  \\
  \midrule
\multicolumn{1}{l|}{Tiny~Images} &
  \multicolumn{1}{l|}{\begin{tabular}[c]{@{}l@{}}Prabhu and Birhane~\cite{prabhu2020large} \\ publish a critique in 2020, \\ showing the dataset \\ contains harmful slurs and \\ many offensive images.\end{tabular}} &
  \multicolumn{1}{l|}{\begin{tabular}[c]{@{}l@{}} Dataset authors formally \\ withdraw the dataset in response, \\  saying images are too small to \\ identify   and remove offensive \\ content and that the dataset’s biased \\ content  violates their values~\cite{torralba2020}.\end{tabular}} &
  \begin{tabular}[c]{@{}l@{}}Dataset taken offline in 2020 and  \\ formally withdrawn via  \\ a letter published on the \\ dataset’s webpage~\cite{torralba2020}. The \\ authors ask the community to \\ refrain from using it in the  future \\  and delete any  downloaded copies.
  \end{tabular} \\  \midrule

  \multicolumn{1}{l|}{\begin{tabular}[c]{@{}l@{}} MS-Celeb \\ -1M \end{tabular}} &
  \multicolumn{1}{l|}{\begin{tabular}[c]{@{}l@{}}MegaPixels investigation in 2019\\ \cite{megapixels,murgia2019s} reveals that the usage \\ of images of “celebrities” \\  are actually private figures and \\ images were scraped  from  \\ the  Internet without consent.\end{tabular}} &
  \multicolumn{1}{l|}{\begin{tabular}[c]{@{}l@{}}No explanation for deprecation \\ is provided on dataset website. \\ Microsoft tells \textit{Financial Times} \\ "the research challenge is over"~\cite{murgia2019microsoft}.
\end{tabular}} &
  \begin{tabular}[c]{@{}l@{}}Dataset deleted from website \\ by Microsoft in 2019; the challenge \\ page has not been updated to \\ mention that the dataset has \\ been deprecated~\cite{msceleb2016}.\end{tabular} \\ \midrule
\multicolumn{1}{l|}{\begin{tabular}[c]{@{}l@{}} Duke  \\ MTMC \end{tabular}} &
  \multicolumn{1}{l|}{\begin{tabular}[c]{@{}l@{}}MegaPixels investigation\\ \cite{megapixels,murgia2019s} spurs Duke University\\ to internally investigate the \\ dataset. This reveals that the  \\  creators violated IRB \\ requirements~\cite{satisky2019duke}.\end{tabular}} &
  \multicolumn{1}{l|}{\begin{tabular}[c]{@{}l@{}}No explanation for deprecation \\ is provided on dataset website. \\ Clarification/apology published by \\ lead author in Duke  student \\ newspaper \textit{The Chronicle}~\cite{tomasi2019}.\end{tabular}} &
  \begin{tabular}[c]{@{}l@{}} Dataset deleted by authors \\ in 2019;  website cannot \\ be reached. \end{tabular} \\
\midrule
\multicolumn{1}{l|}{MegaFace} &   \multicolumn{1}{l|}{\begin{tabular}[c]{@{}l@{}}A \textit{New York Times} investigation \\ suggests use of Flickr images \\ violates Illinois Biometric \\ Information  Privacy  Act~\cite{hill2019photos}.\end{tabular}} & \multicolumn{1}{l|}{\begin{tabular}[c]{@{}l@{}} The dataset website states that the \\ challenge’s goals have been met \\ and ongoing maintenance would be \\ an administrative burden, so the \\ dataset  was decommissioned~\cite{megaface}. \end{tabular}} &
  {\begin{tabular}[c]{@{}l@{}} Statement on website says \\ dataset  has been \\ decommissioned  and  is no \\ longer being distributed.\end{tabular}} \\  \midrule
 \multicolumn{1}{l|}{\begin{tabular}[c]{@{}l@{}} HRT \\ Transgender
\end{tabular}} & \multicolumn{1}{l|}{\begin{tabular}[c]{@{}l@{}} Article on dataset in \textit{The Verge} \\  draws attention to YouTube data \\ scraping issues and lack of \\ YouTube video creator consent  \\ in data  collection~\cite{vincent2017transgender}.
\end{tabular}}  & \multicolumn{1}{l|}{\begin{tabular}[c]{@{}l@{}} Lead researcher tells \textit{The Verge} \\ that the  dataset only \\  contained links to videos,  not the \\ videos themselves; it was never \\ shared commercially, and he \\ stopped  giving access to it \\  three years prior~\cite{vincent2017transgender}. \end{tabular}} &
  {\begin{tabular}[c]{@{}l@{}} Dataset not linked to on \\ research   group’s website, \\ and the URL returns \\ a 404 error.\end{tabular}} \\
  \bottomrule
\end{tabular}
\caption{Examples of deprecated datasets, the external critiques they attracted, and the actions taken by dataset creators.}
\label{table1}
\end{table*}
%\vspace*{-0.2cm}%
There are several commonalities among the removal processes of each datasets, and we have identified five major categories of concern: 
 
\paragraph{Inconsistent public notice and transparency regarding rationale for withdrawal.} While ImageNet, Tiny Images, and MegaFace gave some public notice of deprecation and remediation on their websites, the MS-Celeb-1M, Duke MTMC, and HRT Transgender cases provided no explanations for the removals; the datasets simply vanished from their websites.  The MegaFace website offers that the dataset was decommissioned due to its goals being met and cites the administrative burden of maintenance.  This explanation, though, appears to have been given shortly after a \textit{New York Times} investigation suggested that the dataset may violate privacy laws~\cite{hill2019photos}. Meanwhile, a lead author of Duke MTMC published a clarification and apology in Duke’s student newspaper \textit{The Chronicle} explaining that the dataset was not collected for the purposes of recognizing individuals, but had violated IRB protocols~\cite{tomasi2019}. With MS-Celeb-1M, following the privacy concerns mentioned in Table~\ref{table1}, Microsoft told the \textit{Financial Times} that the dataset had been removed not because of privacy concerns, but because the “research challenge is over.”~\cite{murgia2019microsoft} However, no explanation is given on the page that originally hosted the data~\cite{msceleb2020}. 

\paragraph{Lack of explicit instruction not to use deprecated datasets.} Tiny Images’ authors offer guidance to researchers, asking them not to use the dataset in future and to delete downloaded versions. No such public guidance is provided by the authors of other partially or fully deprecated datasets. This makes it difficult for ML researchers to know if or why they should stop using a dataset in their work, and for reviewers to know whether a dataset used to train a model is acceptable, or if its creators have deprecated it. It also creates uncertainties about the status of derivatives produced from a deprecated dataset, as it is unclear if the reasons for deprecation also apply to the derivative datasets, or if the creators resolved the issues that originally motivated the deprecation. 

\paragraph{Continued circulation of deprecated datasets.} Despite the removal of a dataset from its original hosting location, deprecated datasets often continue to circulate, are used to train models, and are cited in ML papers. Sometimes this research is published years after the deprecation. For example, MS-Celeb-1M’s harms and deprecated status were well-documented in popular press accounts (e.g.~\cite{brandom2019microsoft,murgia2019microsoft,pearson2019}) when it was removed in April 2019. It is also noted as deprecated in the Papers With Code repository, a retraction that proliferates across other dataset collections, like \href{https://datasetsearch.research.google.com/search?query=MS-Celeb-1M\%3A\%20\%7BA\%7D\%20Dataset\%20and\%20Benchmark\%20for\%20Large-Scale\%20Face\%20Recognition\&docid=L2cvMTFtaHJtdHh2cw\%3D\%3D}{Google Dataset Search}. Yet, as Peng et al. have noted, the underlying data for MS-Celeb-1M were used hundreds of times in published papers since its 2019 retraction~\cite{peng2021mitigating}. Today, it continues to circulate on sites like Academic Torrents~\cite{msceleb2019, peng2021mitigating}. In fact, the dataset was uploaded there less than two months after its retraction by Microsoft~\cite{msceleb2019}. MS-Celeb-1M is not the only deprecated dataset with a post-deprecation afterlife.  \href{https://github.com/sxzrt/DukeMTMC-reID\_evaluation}{Duke MTMC}, \href{https://exposing.ai/megaface/}{MegaFace} and \href{https://academictorrents.com/details/03b779ffefa8efc30c2153f3330bb495bdc3e034}{Tiny Images} have also been deprecated but are still widely circulated, used, and cited in both industry and academia (e.g ~\cite{braso2020learning,zhong2020random,li2021head,liu2021head}). Tracking down and auditing all the existing locations of these datasets is a challenge in itself, given the decentralized nature of data hosting and sharing. 

\paragraph{Lack of central directory of deprecated datasets.} Our analysis has focused on the deprecation of highly visible and controversial datasets, but not all deprecated datasets receive such public scrutiny. Popular repositories like Papers With Code and Exposing.ai (the successor to the MegaPixels project by Adam Harvey and Jules LaPlace~\cite{harvey2019exposing}) are useful ways to track datasets that have already been investigated and deprecated, but there remains a need for a centralized resource that aggregates the deprecations of datasets that the ML community has used, regardless of the dataset's prominence or role in a public controversy. To understand which datasets they should use -- and why -- researchers need to be able to see the status of active, updated, and deprecated datasets in one place.  Such a directory would help researchers make informed choices about model training and could, over time, show the ML community how and why it deprecates datasets and become a valuable tool for reflecting on the field's values and practices.

\paragraph{No systematic re-evaluation of related datasets.} Critical accounts ~\cite{crawford2019excavating,yang2020towards, prabhu2020large} have highlighted the challenges of image datasets that rely on the WordNet taxonomy, which contains many offensive and harmful categories. Yet while ImageNet and Tiny Images have been remediated or withdrawn, other datasets based on them, such as CIFAR-10 and CIFAR-100~\cite{krizhevsky2009learning} and Tencent ML-Images~\cite{wu2019tencent} remain active. Meanwhile, some datasets sourced from Flickr, such as MegaFace, may potentially violate Creative Commons licensing and run afoul of privacy laws. But such datasets are only audited on an ad-hoc basis. The decision to deprecate is left to the discretion of dataset creators and their institutions in ways that may idiosyncratically perpetuate harms. \\

Datasets such as the ones listed in Table~\ref{table1} are highly visible and powerfully illustrate some of the challenges of dataset deprecation. Some of them have recently been discussed in Peng et al.~\cite{peng2021mitigating}, who identified \textit{“lack of specificity and clarity”} as to why and how datasets are retracted and observed that even when a dataset is deprecated, copies and derivatives often persist~\cite{peng2021mitigating}. We support this observation, and show in our analysis that the process of deprecation varied greatly amongst the datasets, with little consistency among what triggered their deprecation, how their removals were publicized, and how many of the technical details of deprecation were made transparent. But what are the consequences of improper dataset deprecation, or the lack of a standardized framework for how to remove, delete, or announce derivatives of a dataset? We explore the issues that can arise from improper deprecation in the next section.

\section{Reasons for Dataset Deprecation}\label{problem}

Dataset deprecation is a necessary part of creating and maintaining datasets. Below we address the technical, legal, ethical and organizational considerations that can call for a dataset to be deprecated, and discuss the consequences of the continued use of problematic datasets.

\subsection{Technical Considerations}

There are several technical reasons why a dataset is deprecated, which can result in downstream issues in model relevance and performance. For instance, one issue is that datasets are static -- i.e. they represent a sample of the world that is frozen in time at the moment the dataset is created. The worlds surrounding datasets change and datasets can quickly become unrepresentative of what they were meant to describe. For instance, a corpus like WordNet, which originated in the 1980s, does not include words like “smartphone” or “Internet,” which are relevant in 2022, but does include words like "washwoman" and "chimneysweep," which are no longer part of mainstream modern discourse. A language model trained on WordNet can therefore fail to perform adequately in tasks ranging from question answering to dialogue because the model is built on data that no longer represents current language.

Several authors have studied the problems of semantic drift in WordNet and other training sets~\cite{bakay2019problems, rohrdantz2011towards,montariol2021models} and proposed ways to address and mitigate them from a technical perspective~\cite{hamilton2016diachronic,kutuzov2018diachronic}.  These approaches, though, are rarely incorporated into modern NLP models, which can be trained on large corpora reflecting decades of language data, including books from previous centuries~\cite{bandy2021addressing}. A dataset such as WordNet could, at least, come with a disclaimer explaining its origin and the temporal shifts that occur in language (given that it could still be used for historical research); at best, the current version of WordNet could be delineated for historical research with a more recent derivative created for use in contemporary systems. Addressing `documentation debt'~\cite{bandy2021addressing} and creating datasheets~\cite{gebru2018datasheets} to promote transparency can spur dataset users to make better informed choices about the relevance of a dataset, even if it has not been deprecated. 

Other reasons for deprecating or updating datasets stem from dataset contamination -- i.e. the presence of the same data in both the training and testing of a given dataset. This is increasingly problematic for datasets scraped from the Internet, which may contain millions of documents or images. In a recent study, Dodge et al.~\cite{dodge2021documenting} analyzed C4, the "Colossal Clean Crawled Corpus" used for training many language models~\cite{raffel2019exploring} and found that up to 14.4\% of its test examples for different linguistic tasks were found verbatim in the training data -- which makes the possibility of rote memorization and therefore ``improved performance" a possibility. A similar study of the incredibly popular CIFAR-10 and CIFAR-100 datasets~\cite{krizhevsky2009learning} found that up to 10\% of the images from the test set have duplicates in the training set~\cite{barz2020we}. The authors of the study proceeded to replace the duplicates they found with new images, creating a \href{https://cvjena.github.io/cifair/}{``fair CIFAR" (ciFAIR) dataset}. However, the original dataset continues to be used by ML researchers despite its contamination, due to decentralized hosting and a lack of communication within the ML community about dataset deprecation. When it comes to filtering large text datasets scraped from the Web, given their sheer size (C4 represents 2.3 TB of data, whereas the Common Crawl has 139TB), filtering them is complex and time-consuming, although approaches have been proposed for reducing duplicates and train-test overlap~\cite{lee2021deduplicating}. In practice, documenting and deprecating these datasets is akin to a game of whack-a-mole, since new versions of the Common Crawl come out every few months. Analyzing what they contain and their degrees of contamination through common evaluation tasks would take significant effort.  
%Other work found attempted to remove training set examples that had significant overlap with testing examples, and found that up to 90\% of the evaluating tasks had overlap in the testing data~\cite{brown2020language}. 

Finally, while datasets can never represent a full complexity of social realities, datasets are central to training ML models of social phenomena; it is thus crucial to trace their origins, assumptions, and representativeness, and to detect failures and deprecate datasets accordingly. For instance, several approaches have been proposed in recent years for `machine unlearning'~\footnote{The term "machine unlearning" is also the name of a working group led by Wendy Hui Kyong Chun and Kate Crawford which critically engages with the way machine learning is understood and generates new approaches to models of learning and validation.}, allowing data to be erased from already trained models~\cite{cao2015towards, baumhauer2020machine, bourtoule2021machine,sommer2020towards}. Other work has also proposed that: 1) it is possible to reveal details from an initial dataset even when a model was subsequently retrained on a redacted version~\cite{zanella2020analyzing}; 2) adversarial querying of a trained language model can recover individual training examples~\cite{carlini2020extracting} and 3) "radioactive" data tracing is possible via imperceptible changes that make it possible to find the origins of data points~\cite{sablayrolles2020radioactive}. While the broader impacts of derivative datasets are to be determined, researchers often underestimate the influence of individual data points and overestimate the effect of fine-tuning on pretrained models. We see this as a further technical reason to support proper dataset documentation and deprecation, to avoid both malicious and inadvertent usage of data that is no longer relevant or that is actively harmful.

\subsection{Legal Considerations}

For most dataset creators, providers, and downstream users, there are also important legal considerations when assessing the appropriateness of deprecation. This includes potential violations of laws governing privacy, discrimination, data protection, intellectual property licenses, fair decision-making processes, consumer protection, and use of an individual's image or likeness, among numerous others. For example, in the United States, numerous legal actions have recently been brought against companies such as Clearview AI and IBM for their facial recognition datasets. Clearview has been sued under California law for commercial appropriation of individual faces within photographs, violations of California’s constitutional privacy protections, and for aiding and abetting illegal government surveillance efforts ~\cite{renderos2021}. Another California lawsuit claims Clearview violated the new California Consumer Privacy Act~\cite{pascu2020}. It has also been sued under Illinois’ Biometric Information Privacy Act (BIPA), where the legal complaint alleges that Clearview illegally “harvested” biometric facial scans from scraped photographs and then also illegally distributed this data to its law enforcement clients~\cite{thornley2021}. Outside the US, Clearview faces investigations and legal claims in the UK, EU, and Australia under various data protection laws, including the General Data Protection Regulation (GDPR)~\cite{ico2020,meyer2021}. Similarly, IBM’s Diversity in Faces (DiF) dataset has been the subject of litigation, including a lawsuit focused on violations of BIPA~\cite{rizzi2020}. Facebook has also settled a case alleging BIPA violations for \$550 million in 2020~\cite{coldewey2020}. Claims of copyright infringement have also been brought against image dataset providers; however, they have yet to succeed, especially in the United States where various fair use legal decisions have created broad permissions for mass digitization and analysis of creative works for machine learning purposes~\cite{2015authors,2007perfect,2009av, levendowski2018copyright,burk2019algorithmic}. 

The landscape of potential legal issues applicable to datasets is complex and will vary based on content, jurisdiction, and application~(e.g.~\cite{isight}). Therefore, it will be important for dataset creators and providers to consistently assess them over time, especially during periods of dramatic changes in the law, such as when GDPR was implemented in May 2018, when the US-EU Privacy Shield replaced the data protection Safe Harbor Framework, or if, for example, the EU AI Act becomes law~\cite{schwartz2019structuring,veale2021demystifying}. Jurisdictional challenges also raise questions as to how deprecation would work across different jurisdictions, with different legal regulations and restrictions. This is an important consideration, which we address below by suggesting the proposed centralized repository note jurisdictional differences and, if applicable, support the use of geo-fence techniques to control access to the dataset going forward.
 
When legal actions involving datasets are successful, the result often impacts the dataset itself and even the ongoing influence that the data has on algorithmic systems. For example, in a recent Federal Trade Commission settlement, Ever Inc. was found to have deceived its customers when it collected their photos, failing to disclose how it used those photos to train facial recognition systems. As part of the remedy for these violations, the settlement requires the company “to delete the photos and videos of Ever app users who deactivated their accounts and the models and algorithms it developed by using the photos and videos uploaded by its users”~\cite{ftc2021}. As these and other legal issues continue to arise involving datasets, providers should monitor them and take them into consideration in their deprecation decisions. This will also impact downstream users, who may incur liability if they use deprecated datasets with legal issues, even unwittingly~\cite{rizzi2020}. This further supports the urgent need for a deprecation framework, as well as the existence of an easily-accessible public repository so researchers can check which datasets present risks and may even cause legal liability if they use them. If a company makes a public commitment to follow the Framework, that promise can generally be legally enforced both by the FTC and State Attorneys General under Truth-In-Advertising/Deceptive and Unfair Practices statutes as well as by the SEC if it is a publicly traded company and the promise is part of their quarterly or annual SEC filing. It is worth noting, however, that while one goal of the public repository is to promote transparency and understanding of the reasons for the deprecation, concerns over admissions of legal liability may limit or inhibit listing all of the specific legal analysis underlying the deprecation decision. Therefore, allowing for some flexibility in the specificity of legal reasons outlined in a deprecation report is advisable.

\subsection{Social Considerations} %maybe we can come up with a better name here, @Kate? Not sure the FAcct community will like the usage of 'ethical' here

Datasets necessarily represent a worldview, including how the data is collected, labelled and applied, and these particular representations and classifications of the world have social and political values embedded within them~\cite{crawford2021atlas, bowker2000sorting}. A growing body of literature within ML has begun to consider the broader social implications of machine learning and artificial intelligence deployment. Such research describes complex ethical, social, and political considerations with which the ML community and field at large are beginning to grapple~\cite{greene2019better,stark2019data, hoffmann2020terms}. For the datasets that we reviewed, the reasons provided for their deprecation were often linked to concerns ranging from consent, privacy, and offensive content to the violation of values like inclusivity and representation~\cite{gitelman2013raw,kitchin2021data}.
Systems shaped by these datasets can produce different forms of harm, including allocative harms (a system offering or withholding opportunities from certain groups) and representational harms (a system reinforcing the subordination of particular groups by virtue of identity)~\cite{barocas2017problem}. As numerous audits of datasets show, such harms tend to disproportionately affect marginalized groups along the intersecting axes of race, ethnicity, gender, ability, and positionality in global hierarchies (see, for instance, \cite{shankar2017no,buolamwini2018gender,crawford2019excavating,hutchinson2020social,prabhu2020large}). 

Research on the social considerations of datasets tends to address dataset creation, and to a limited degree, dataset maintenance~\cite{peng2021mitigating}. While other scientific domains have addressed the harms of redacted papers that continue to circulate~\cite{grieneisen2012comprehensive,bar2018temporal}, rarely do scholars consider the deprecation and post-deprecation phases of a dataset life cycle. Yet the deprecation phase is crucial when evaluating the wider implications of datasets, because deprecation is often done in response to perceptions, accounts, or critiques of a dataset’s harmful impacts. For instance, in a rare case of explicit explanation about dataset deprecation, the creators of Tiny Images noted that the repository was taken offline because it contained “biases, offensive and prejudicial images, and derogatory terminology,” in part due to automated data collection from WordNet~\cite{miller1990introduction}, which threatened to “alienate an important part of our community”~\cite{torralba2020}. The deprecation notice for Tiny Images was posted in direct response to a critique by external researchers, who showed that the dataset contained racist and misogynist slurs and other offensive terms, including labels such as “rape suspect” and “child molester”~\cite{prabhu2020large}. These labels are attached to images of people downloaded from the internet who did not give their consent, constituting a clear form of representational harm and potential defamation~\cite{barocas2017problem}.  Similarly, after public and scholarly criticism, the creators of ImageNet~\cite{deng2009imagenet} identified a total of 1,593 harmful labels in the dataset, and subsequently removed them~\cite{yang2020towards}. However, because both Tiny Images and unredacted versions of ImageNet continue to circulate and are potentially used to train production-level systems, these problematic labels and logics could be embedded in ways that entrench harms while being hard to track and investigate: for instance, if an image classification model trained on ImageNet is used for predicting criminality based on nothing except an individual’s perceived similarity to that of a ‘rape suspect.’ The continued circulation of “toxic” data collections therefore threatens to reproduce both allocative harms and representational harms. 

The deprecation of datasets can also change the dynamics of harm. When datasets are deprecated by their creators without disclosure or discussion of the types of harms they reproduce, this can create widespread uncertainty. Those responsible for the dataset have not acknowledged the potential for harm, while simultaneously relinquishing any control over the afterlife of the dataset. Rather than mitigating potential harms at the end of the dataset life cycle, the creators have merely removed a single node for accessing the dataset while perpetuating a view of the data’s supposed appropriateness for training models. Moreover, when deprecated datasets continue to be accepted and cited in ML conferences, the broader field implicitly condones their continuing circulation and possible harms.

\subsection{Organizational Considerations}

While technical, legal, and ethical considerations are of primary importance to understanding the downstream effects of how deprecated datasets continue to circulate, organizational concerns also deserve attention because of the role that varied entities play in the stewardship of training data. As shown in Table~\ref{table1}, datasets are created and maintained by both industry research groups like Microsoft (e.g. MS-Celeb-1M) as well as by academic groups like the Stanford Vision Lab (e.g. ImageNet). While maintenance work is often devalued and rendered invisible~\cite{russell2018after}, maintaining these datasets over time requires organizations to invest significant resources, including human labor, technical infrastructure, and financial support. Organizational changes often necessitate the removal of data~\cite{corry2021a}, because of varied factors -- from a shifting research team to the end of grant funding -- may spell a decreased ability to maintain the dataset properly. In these cases, it may be better to deprecate a dataset conscientiously than to maintain it poorly.

As we have shown in Section~\ref{outline}, however, dataset deprecation is often done hastily, spottily, and without adequate documentation. These faulty dataset deprecations can have potential downstream effects for the organizations in question, particularly in terms of reputation management. Scholars of organizational communication have shown that data maintenance -- in the form of data breaches/data security -- are important factors for organizational reputation today~\cite{bentley2020testing,corradini2020data}, and researchers are proposing new forms of data management and stewardship that acknowledge organizational complexities~\cite{bennett2021pathways}. As people increasingly become aware of the implications of ML datasets through media coverage (e.g. ~\cite{murgia2019microsoft}), and even of the implications of the ongoing use of problematic datasets~\cite{hao2021}, it is in organizations' best interests to steward datasets properly through the end of their lives. Moreover, this stewardship will need to be domain-specific and tailored to the needs and cultures of different organizations, professional communities, and historical traditions. 

The reasons listed above illustrate the need for a common deprecation framework for making this process more transparent and coherent, while helping both dataset creators and downstream users address problems and concerns, improve communication, and establish the trust necessary for proper dataset stewardship through all parts of the data life cycle. In the following section, we propose a general framework designed in consultation with ML researchers and practitioners to equip dataset creators, users, and other members of the community with tools for responsible dataset deprecation.

\section{Dataset Deprecation Framework} \label{framework}

To further the conversation about the governance of dataset deprecation and build on existing methods, we propose a Dataset Deprecation Framework that could be adopted by dataset creators, consisting of three elements: a dataset deprecation report, a unique identifier, and a centralized repository for maintaining status updates.

\subsection{Dataset Deprecation Report}
%Part of dataset governance is dataset deprecation and a greater accountability for the afterlife of a dataset. 

The Dataset Deprecation Report is the central element of our proposal, consisting of six key elements influencing dataset deprecation and the necessary information to communicate to the ML community the procedures to follow during and after dataset deprecation. We provide one concrete example of how this framework can be implemented by dataset creators in Table~\ref{example} below, and an additional one in the Appendix.

\begin{enumerate}
    \item \textbf{Reasons for Deprecation}: Using publicly accessible explanations, those responsible for a dataset should clearly explain potential impacts of the deprecation. These risks should be enumerated with careful attention given to impacted communities that have historically lacked social power and institutional standing in computational and data science research. The discussion of risks could include which risks are being envisaged and to whom, and over what time frame risks were considered (such as identifying short, medium, and long-term risks if relevant). 
    \item \textbf{Execution and Mitigation Plan}: Drawing on the history of how technology companies have approached deprecation of software and other technological tools~\cite{corry2021}, the parties implementing dataset deprecation should provide an Execution Plan regarding how the dataset deprecation will happen, and how any adverse impacts from its use will be mitigated. These should include how access to the dataset will be restricted or halted; how changes to access will be announced and maintained on a publicly accessible site; what steps are being taken to limit or prevent downstream uses; and which derivative datasets are impacted and should potentially also be deprecated.
    \item \textbf{Appeal Mechanism}: An appeal mechanism should be included that allows challenges to the deprecation. Appeals should have clear and well-defined processes, with a mechanism to contact a person responsible for the appeal process, including timely responses and explanations. 
    \item \textbf{Timeline}: Deprecation announcements should give stakeholders adequate time to understand the deprecation’s rationale, evaluate its impact, and launch any appeals. At a minimum, deprecation timelines should take into consideration an analysis of the dataset’s ongoing harms and risks and the perceived impact on stakeholders, with consideration given of contexts that require action to mitigate harms versus those that can allow for more time.
    \item \textbf{Post-Deprecation Protocol}: Recognizing that deprecated datasets will continue to have value (e.g., as research objects, legal evidence, and historical records), deprecation protocols should also articulate methods for sequestering and accessing datasets post-deprecation, including what principles and procedures will be used to grant access to sequestered datasets. This protocol should be regularly re-evaluated in light of technical changes in data sequestration, new best practices for policy implementation, and any insights gained during the appeals process.
    \item \textbf{Publication Check}: Leading technical conferences such as NeurIPS and ICML could include in their paper acceptance protocols a requirement that all authors affirm that their work does not use deprecated datasets, and their work follows the post-deprecation protocols sanctioned by the conference, or risk that their work be rejected. Additionally, those who are presenting new datasets at conferences should note in their paper that they will follow the conference's framework for dataset deprecation, in the event that their datasets require future removal or modifications.
\end{enumerate}

 We recognize that the rationales and motivations for deprecation will vary, especially given domain-specific issues, and that the circumstances triggering deprecations may cause varying degrees of disclosure and explanation. We also recognize the need for the proposed report to be interpreted and adapted to meet the needs of localized domains of knowledge and diverse organizational aims. For example, where deprecation rationales stem from legal considerations, dataset providers may not be in a position to share details fully without risking legal liability. Or, similar to responsible vulnerability disclosure policies, when disclosure could result in additional harm, the timing and specificity of the disclosure may vary as well. Bracketing a full discussion of the legal and normative motivations for dataset deprecation (cf.~\cite{householder2017cert}), and recognizing that each deprecation has contingencies and nuances that go beyond the scope of this paper, we propose these elements as generic enough to be used in most cases of dataset deprecation.

\subsection*{Example 1 -- Dataset Deprecation Report: FaceFeels}

\noindentparagraph{Deprecation Scenario:} A research lab built a large dataset called FaceFeels. It contains six million images of people’s faces scraped from the internet for the purposes of labelling facial expressions and training systems for emotion detection. The full FaceFeels dataset was made available on a website but downloads were restricted to people who applied with a university email address. The lab kept a record of email addresses but did not require any details for what the data would be used for. After growing concerns about privacy and lack of consent from the people in the photographs, as well as the lack of scientific consensus about the validity of inferring emotion from facial expressions, the lab decided to deprecate the dataset. The company completes the Dataset Deprecation Report below and submits it to a centralized repository:

\begin{table*}[ht!]
\begin{tabularx}{15cm}{l|X}
\toprule
 \multicolumn{1}{l|}{\begin{tabular}[c]{@{}l@{}}\textbf{Reasons for} \\ \textbf{Deprecation} \end{tabular}}   & We have deprecated the FaceFeels dataset due to issues raised in the original context for scraping the photographs, methodological concerns about universal emotion detection, and the potential for harm in downstream implementations. We recognize that removing the dataset will potentially have commercial and scholarly impacts; however, we feel the risks outweigh any of those potential benefits and therefore are deprecating the dataset. 
 \\ \midrule 
\multicolumn{1}{l|}{\begin{tabular}[c]{@{}l@{}}\textbf{Execution \&} \\ \textbf{Mitigation Plan} \end{tabular}} & We have attempted to contact all stakeholders who applied to download the dataset  via email to request that they no longer use it. We have sent them this protocol, explaining all aspects of the deprecation plan, and provided them with a single point of contact within the lab who can answer their questions about the deprecation. We recognize that the deprecation will impact stakeholders on different timeframes and will make good faith efforts to work with stakeholders to ensure that the deprecation creates the least amount of disruption possible. 
Additionally, we have located the FaceFeels dataset on the CentralDataPlus repository. We have emailed the administrators of this repository, and asked them to remove the dataset within 30 days and publish the following note:
\textit{“The FaceFeels dataset was removed from this site on <date> at the request of the creators. A complete account of the deprecation is contained in this Dataset Deprecation Report <linked> and further questions about the deprecation can be directed to the FaceFeels team here: <contact information>. Additionally, we are placing a notification of deprecation on the Dataset Deprecation site, with a link to the full protocol.
We have emailed researchers who have downloaded the dataset from our site, informed them of the deprecation, shared a copy of the deprecation protocol, and asked them to stop using the dataset and to add post-publication notifications where possible informing readers of the deprecation.”}
\\ \midrule 
\multicolumn{1}{l|}{\begin{tabular}[c]{@{}l@{}}\textbf{Appeal} \\ \textbf{Mechanism} \end{tabular}} & Appeals will only be allowed in limited cases with a strong justification. We give stakeholders 30 days from the date of outreach to inform us of any requests for ongoing use. Appellants should identify any disruptions caused by the deprecation, and can request modifications (e.g., timeline extensions) if needed. 

We will evaluate the appeal within 15 days of receipt, and communicate our decision via email. We will publicly update the protocol if there are any changes that affect all stakeholders.

Appeals received more than 30 days after deprecation will not be considered.  \\ \midrule
  \multicolumn{1}{l|}{\begin{tabular}[c]{@{}l@{}}\textbf{Timeline} \end{tabular}} & The dataset will be removed after 30 days of the public announcement to deprecate it. It will be removed from our site, and ideally from all repositories (assuming they agree to our request). 
  \\ \midrule
 \multicolumn{1}{l|}{\begin{tabular}[c]{@{}l@{}}\textbf{Post-Deprecation} \\ \textbf{Protocol} \end{tabular}} & Our lab will retain a complete copy of the FaceFeels deprecated dataset. It will not be available publicly, or to researchers without approved access. Access will only be given to researchers who are doing fairness and auditing work, such as understanding how the dataset is impacting production-level systems. Access will not be approved to train facial recognition or emotion detection tools. Researchers can apply at this email address <contact details> and they must stipulate a research rationale, use strict access protocols, and agree to our terms of access and timeline for use. 
 \\ \midrule 
 \multicolumn{1}{l|}{\begin{tabular}[c]{@{}l@{}}\textbf{Publication} \\ \textbf{Check} \end{tabular}} &  We ask all researchers to stop using the FaceFeels dataset, other than researchers who have been given post-deprecation approval. Researchers submitting conference papers that use the FaceFeels dataset with approval should include with their submissions a copy of the approved post-deprecation protocol.
 \\ \bottomrule 
\end{tabularx}
\caption{An Example of a Dataset Deprecation Report, for the fictional FaceFeels dataset} 
\label{example}
%vspace*{-0.8cm}%
\end{table*}

\subsection{DOI Identifiers}

The protocol described above is a key part of our deprecation framework. But in order to actuate its effect with a strong technical infrastructure, we propose that it be supported by a permanent identifier that can accompany datasets from their creation to their deprecation (as well as through updates and version changes). A suitable mechanism for assigning and updating identifiers could be based on the DOI (digital object identifier) system, which has existed for decades but has been largely overlooked by the ML community~\cite{scheuerman2021datasets}. We start with a brief history of the DOI and outline its potential for streamlining and standardizing the dataset creation, maintenance and deprecation process.

The digital object identifier, or DOI, was introduced by the International Organization for Standardization (ISO) in 2000 as a way to standardize and track the creation and evolution of digital objects ranging from academic articles to government publications. DOIs are fixed, but are bound to metadata -- including the object's URL, date of creation, authors, version, etc. -- that can be changed. These changes are automatically tracked during the object's life cycle. DOIs are governed by the International DOI Foundation (IDF), a non-profit entity that ensures the perpetuity of the format and prevents third parties from imposing licensing on the system. The attribution of DOIs is made by registration agencies, which are appointed by the IDF and provide users with ways to register new DOI objects and maintain and update existing ones. Websites such as \href{https://zenodo.org/}{Zenodo} and \href{https://figshare.com/}{FigShare} have been offering free DOI attribution and maintenance for the academic community for almost a decade, covering various types of outputs ranging from articles to figures and datasets, with the philosophy that \emph{``Research objects need to be citable in order to be usable."}~\cite{figshare}. 

The practice of using DOIs to track the creation and modification of datasets remains a rare occurrence in machine learning. Recent research in computer vision found that only about 3\% of datasets use DOIs, with the overwhelming majority of dataset creators using personal websites for hosting datasets~\cite{scheuerman2021datasets}. This makes it not only hard to find datasets, but to determine their status and current version. Also, while websites such as GitHub, which are also often used for disseminating data, support versioning and change tracking, they often lack structured metadata, datasheets, and inter-operability with similar systems. Although using either personal websites or code-hosting repositories may allow for more flexibility in description and indexing, they lack any kind of traceability or standardized structure. In fact, both of these platforms can be used in conjunction with DOI; i.e., a dataset hosted on a private website can still be assigned a DOI with its accompanying metadata and unique identifier, which will continue to persist as a trace even following their eventual modification or deprecation.

The existence and usage of DOIs as perennial identifiers of digital objects can also serve as a basis for building similar infrastructure specifically tailored for the ML community. For instance, Zenodo permits the creation of `Communities' on its platform by researchers from a given domain or area of expertise; the members of a community can co-curate and co-construct what is accepted or not by the community. Many projects, workshops and journals have their own communities, ranging from the \href{https://zenodo.org/communities/biosyslit}{Biodiversity Literature Repository} to the \href{https://zenodo.org/communities/data-science-in-healthcare/}{Journal of Data Science in Healthcare}. Each object in the communities is assigned its own persistent DOI, along with other relevant metadata such as author name, date, provenance, institution, etc. -- all fields that can be tracked using versioning systems and cited using tools like BibTeX, already used by many. Leveraging permanent identifiers such as DOI and adding comprehensive information about datasets can help our community improve the communication and standardization of datasets. Moreover, replacing the dataset itself with its data deprecation report upon removal can help facilitate and
normalize proper dataset care through the end of the dataset life cycle. By learning from and building on the existing infrastructures and practices of data stewardship, ML researchers and practitioners can substantially improve how datasets are created, maintained, and deprecated. 

\subsection{Central Repository of Deprecated Datasets}

Having a central place to store, update and disseminate dataset deprecation notes is important to ensure that the ML community is aware of dataset deprecations in real time. We propose that a leading conference such as NeurIPS or a collegial entity such as AAAI, IEEE, or ACM act as the keeper of a public, centralized repository of deprecation decisions. This could take the form of a database of deprecation sheets and their accompanying documentation. This would give researchers a single site to visit in order to check if a dataset has been deprecated, thus protecting them from a range of technical, legal, ethical, and organizational problems. It would enable a transparent way to submit, access, and disseminate up-to-date information on a dataset's  current status, as well as notifying the ML community when a deprecation is necessary and why. This would also support jurisdiction-specific deprecations, such as when a dataset is illegal in one jurisdiction but allowed in others. The repository maintainer could address this by noting jurisdictional differences in the repository and, if applicable, supporting geo-fence techniques to control access to the dataset going forward. 

Furthermore, a central repository would enable a reporting function, where people could share issues they have discovered with existing datasets, even if they are not the original creators. Researchers often discover problems with datasets in the process of their work, but currently have limited options for informing others who are using the same data. A centralized repository would be a useful venue to inform the wider community of any potential issues.  It could also become, over time, a record of how the ML community analyzes and acts upon dataset harms.

As standard bearers in the ML research community, major technical conferences hosted by NeurIPS, ICML, and FAccT could implement the deprecation publication check for paper submission that we have proposed here. This is an important step, as paper acceptance protocols are a powerful way to create norms across a research community. In this way, conferences can play a gatekeeping role that contributes to curbing the circulation and use of deprecated datasets. It is important to acknowledge the labor, in terms of time and effort, that would be required to host an up-to-date repository of deprecated datasets. It requires an ongoing institutional commitment to the maintenance of this infrastructure, which is why we see it as a task for prominent, well-established, and well-funded conferences and industry bodies~\cite{jackson2017speed}. This is another reason why our approach places the responsibility of repository creation and maintenance with leading ML institutions, while dataset creators are responsible for completing the Dataset Deprecation Report and lodging it with the repository. This workflow may also represent the opportunity for fruitful collaboration with institutions that have significant experience in research ethics and information management, like a university institutional review board (IRB) or university library. Staff from either or both institutions could be valuable resources for researchers wishing to associate a dataset with a DOI or fill out a Dataset Deprecation Report.

\section{Conclusion}

Dataset deprecation is an important part of the dataset life cycle, but it is yet to receive sufficient attention from the ML community. Developing care practices for properly dealing with deprecation, as well as mechanisms and frameworks to support these practices, are key parts of conscientious data stewardship. In this paper we have proposed a novel \textit{Dataset Deprecation Framework}, which includes three elements: a dataset deprecation report, a system of permanent identification, and a centralized repository. The proposed framework was informed by our review of several major datasets that have been redacted or deprecated, where we observed that existing deprecations have been subject to poor documentation practices, leaving the ML research community with ongoing uncertainty and no clear rationale for stopping their use. The deprecation report first helps address these issues by allowing dataset creators to document the reasons for removing a dataset and outline an execution and mitigation plan, an appeal mechanism, a timeline, and a post-deprecation protocol. The continuing circulation of these datasets poses significant technical, legal, and ethical problems, as well as domain-specific issues. In combination with dataset care practices like datasheets, checklists, and audits, the deprecation framework offered here provides a way for ML communities to practice responsible and ethical data management throughout a dataset’s lifecycle. 
 
Deprecation reports are only one part of the deprecation process -- greater infrastructure is required to mitigate their use. In addition to reports, we have proposed the idea of a central repository where all deprecated datasets are listed, an identification mechanism that can help identify datasets in a permanent way, as well as a publication check for major conferences to ensure that emerging research no longer uses these data sources. We recognize that the maintenance of a centralized deprecation database is an administrative challenge, but it is also an important one; in addressing this to the ML research community, the aim is to support a growing culture of communal, care-based practices, calling for greater attentiveness to the labor and impact of producing, maintaining, and sunsetting datasets. In general, our proposal of a deprecation framework leads toward greater field-level considerations around the infrastructures (or lack thereof) that are available for the ethical management of ML datasets. While we have proposed tools for use in deprecation, future conversations about data stewardship might consider the increased participation of existing ethics bodies like university IRBs or corporate ethics boards during dataset creation, use, and indeed deprecation. 

Crucially, however, this work highlights what is inevitably a broader problem that proliferates in less transparent environments, including industry and personal contexts. For instance, not all deprecated datasets are stored on public online repositories, as datasets can be stored privately by an individual or company in perpetuity. Moreover, deprecated datasets may be used in industrial applications, where models trained on these data are deployed in technological systems but leave no citation record for outside accountability. Additionally, the continued use of datasets with known problems highlights the urgent need to attend to the production of derivative datasets from deprecated ones, and the challenge of tracing any inherited issues, a topic the authors are addressing in forthcoming work. By creating a deprecation framework and proposing infrastructures to support the full data lifecycle, we hope to underscore the importance of a community-based approach to care practices for datasets, while raising awareness about the larger issue of the afterlives of deprecated datasets in academia, industry and beyond.

\begin{acks}
The authors wish to thank the industry practitioners and academic researchers who gave feedback on the framework outlined here, with particular gratitude to Adam Harvey for his detailed insights. We also want to acknowledge the many  conversations about better dataset stewardship with members of the FATE research group at Microsoft Research, including Solon Barocas, Hal Daumé III, Miroslav Dudik, Michael Madaio, Ida Momennejad, Alexandra Olteanu, Jennifer Wortman Vaughan, Hanna Wallach, and  Josh Greenberg. We want to acknowledge the support of the Alfred. P Sloan Foundation, as part of their funding of the Knowing Machines project.
\end{acks}
\newpage 

\bibliography{bibliography}
\bibliographystyle{ACM-Reference-Format}
\newpage
\section*{Appendix}

\section*{Example 2  Dataset Deprecation Report: DataDriver}
Using a hypothetical dataset and deprecation motivation, this example report offers another illustration how the dataset deprecation process could be implemented.

\paragraph{Scenario:} The car insurance company DataDriver built a large dataset of driving behaviors and contexts (e.g., routes taken, mileage driven, driving speeds, maintenance alerts, weather conditions, driver preferences, etc.) and made this full dataset available for download through a variety of community dataset sharing sites. The company did not closely track downloads, did not distribute the dataset through an API, and periodically uploaded new versions of the dataset. For a variety of reasons, DataDriver has now decided to deprecate the dataset. Not knowing exactly where this dataset may appear, but knowing that it may be part of a variety of commercial, noncommercial, industrial, and research systems, the company wants to follow the Dataset Deprecation Framework. The company completes the Dataset Deprecation Report below and submits it to the centralized repository.

\begin{table*}[ht!]
\begin{tabularx}{15cm}{l|X}
\toprule
 \multicolumn{1}{l|}{\begin{tabular}[c]{@{}l@{}}\textbf{Reasons for} \\ \textbf{Deprecation} \end{tabular}}   & After reviewing the dataset, DataDriver realized that the dataset had gradually developed links to multiple data sources. Although its dataset contained no personal information (e.g., driver names, addresses, or license plate numbers), DataDriver was concerned that, when this dataset is combined with other datasets (e.g., those showing driving routes, traffic patterns, or travel times), users could be de-anonymized in ways that the company had not originally expected. DataDriver is deprecating the dataset out of abundance of caution, to protect against any unintentional identity disclosures. 
 
 Deprecating DataDriver creates risks to researchers who rely on the dataset for ongoing scholarly inquiry. Stakeholders may include researchers studying a variety of contexts—e.g., human driving behavior, ecological impacts of car travel, vehicle maintenance, safety procedures and collision conditions, and more. We recognize that removing the dataset will impact scholars’ ability to design longitudinal studies and do comparative research. Additionally, stakeholders may include commercial developers doing market and product research for new businesses, service designs, and industry innovations. Deprecating the dataset may impact their market and product knowledge. Finally, the dataset may be informing public and private sector research on the car industry as a whole, so we recognize that deprecating the dataset may interrupt their knowledge of the automotive field, regulatory initiatives, and market trends.
 
 Tracing the DataDriver dataset on various dataset repositories (e.g., XXXXX), we have made considerable efforts to contact stakeholders we are aware of who have relied on the dataset and may continue to do so. We have sent them this protocol, explaining aspects of the deprecation plan, and provided them with a single point of contact within the company who can answer their questions about the deprecation. We recognize that the deprecation will impact stakeholders on different timeframes and will make good faith efforts to work with stakeholders to ensure that the deprecation creates the least disruption possible. \\ \midrule 
\multicolumn{1}{l|}{\begin{tabular}[c]{@{}l@{}}\textbf{Execution \&} \\ \textbf{Mitigation Plan} \end{tabular}} & We have identified the DataDriver dataset on the following dataset repositories: X, Y, Z. We have contacted the administrators of each repository, identified ourselves as the originators of the dataset, and have requested that they remove the dataset from their sites within 30 days. Additionally, we have requested that they not only remove the dataset from their repository but that they also retain a repository entry with the following note: \begin{quote}
   \textit{The DataDriver dataset was removed from this repository on <date> at the request of the DataDriver company. A complete account of the deprecation is contained in this Dataset Deprecation Report <linked> and further questions about the deprecation can be directed to DataDriver <contact information>. Additionally, we are placing a notification of deprecation on the Dataset Deprecation site, with a link to the full protocol.}
\end{quote}

We have also made our best efforts to identify the authors of reports and papers that use the DataDriver dataset and have written to those individuals informing them of the deprecation, sharing a copy of the deprecation protocol, and asking them to add post-publication notifications where possible informing report/paper readers of the deprecation.
\\ \midrule
 \multicolumn{1}{l|}{\begin{tabular}[c]{@{}l@{}}\textbf{Appeal} \\ \textbf{Mechanism} \end{tabular}} & In each outreach to dataset stakeholders, we specified that stakeholders have 30 days from the date of outreach to appeal the deprecation. Appellants should clearly articulate the disruptions to them caused by dataset deprecation. Additionally, appellants can request changes to the deprecation process (e.g., timeline extensions) and propose any alterations to the protocol that they think would reduce deprecation disruptions.

We commit to evaluating the appeal within 30 days of receipt, communicating our decision to appellants, and changing the protocol / deprecation execution plan, if appropriate.  We will publicly update the protocol with any changes and inform all stakeholders of the modifications.

Appeals received more than 30 days after deprecation will only be considered in extreme cases.
 \\ \midrule 
  \multicolumn{1}{l|}{\begin{tabular}[c]{@{}l@{}}\textbf{Timeline} \end{tabular}} & Within 30 days of the deprecation’s public announcement and communication to stakeholders, the dataset will be removed both from our website and the dataset repositories (subject to those repositories’ responsiveness).
 \\ \midrule 
 \multicolumn{1}{l|}{\begin{tabular}[c]{@{}l@{}}\textbf{Post-Deprecation} \\ \textbf{Protocol} \end{tabular}} & We will retain, internally within our company and not through any publicly available repository, a complete copy of the deprecated dataset.

Researchers wanting access to the deprecated DataDriver dataset can apply to us <contact information> for access. Applicants must state a research rationale, complete our access training module that specifies access protocols, and agree to our terms of access. Researchers will not be given a copy of the dataset but will instead access/query the dataset through our process for accessing sequestered datasets. \\ \midrule
\end{tabularx}
\end{table*}
\newpage
\begin{table*}[ht!]
\begin{tabularx}{15cm}{l|X}

 \multicolumn{1}{l|}{\begin{tabular}[c]{@{}l@{}}\textbf{Publication} \\ \textbf{Check} \end{tabular}} &  We are submitting this protocol to the relevant Dataset Deprecation Site and ask all researchers submitting papers to NeurIPS to confirm that their submissions do not use the DataDriver dataset, with the exception of researchers who have followed our post-deprecation protocol. Researchers submitting NeurIPS papers that use the DataDriver dataset should include with their submissions a copy of the approved post-deprecation protocol.
 \\ \bottomrule 
\end{tabularx}
\caption{A Second Example of a Dataset Deprecation Report, for the fictional DataDriver dataset} 
\end{table*}
\clearpage

\end{document}